\documentclass[a4paper, pra, twocolumn, superscriptaddress,showpacs]{revtex4}

\usepackage{amssymb}
\usepackage{amsmath}
\usepackage{color}
\usepackage{graphics, graphicx}
\usepackage{psfrag}
\usepackage{mathcomp}
\usepackage{subfigure}
\usepackage{color}
\usepackage{soul}

\begin{document}

\title{Pressure induced magnetic phase separation in La$_{0.75}$Ca$_{0.25}$MnO$_{3}$ manganite.}

\author{M.~Baldini}
\affiliation{Dipartimento di Fisica, Universit\`a ``Sapienza",
P.le A. Moro 4, 00187 Roma, Italy}
\affiliation{HPSynC, Geophysical Laboratory, Carnegie Institution of Washington, 9700 S. Cass Avenue, Argonne, IL 60439,
USA}
\author{L. ~Capogna}
\affiliation{Consiglio Nazionale delle ricerche IOM-OGG c/o Institut Laue Langevin 6 rue
J. Horowitz 38042 Grenoble, France}
\author{M.~Capone}
\affiliation{Consiglio Nazionale delle Ricerche, Istituto Officina dei
Materiali (CNR/IOM), Uos Democritos, SISSA, International School
for Advanced Studies (SISSA/ISAS), Via
Bonomea 265, 34136 Trieste, Italy}
\affiliation{Dipartimento di Fisica, Universit\`a ``Sapienza",
P.le A. Moro 4, 00187 Roma, Italy}
\author{E.~Arcangeletti}
\affiliation{Dipartimento di Fisica, Universit\`a ``Sapienza",
P.le A. Moro 4, 00187 Roma, Italy}
\author{C.~Petrillo}
\affiliation{Dipartimento di Fisica, Universit\`a di Perugia,
06123 Perugia, Italy}
\affiliation{Consiglio Nazionale delle Ricerche, Istituto Officina dei
Materiali (CNR/IOM), Uos Democritos, SISSA, International School
for Advanced Studies (SISSA/ISAS), Via
Bonomea 265, 34136 Trieste, Italy}
\author{I.~Goncharenko}
\affiliation{Laboratoire L\'eon Brillouin, CEA-CNRS, CE-Saclay, 9191
Gif-sur-Yvette, France}
\author{P.~Postorino}
\affiliation{Dipartimento di Fisica, Universit\`a ``Sapienza",
P.le A. Moro 4, 00187 Roma, Italy}
\affiliation{Consiglio Nazionale delle Ricerche, Istituto Officina dei
Materiali (CNR/IOM), Uos Democritos, SISSA, International School
for Advanced Studies (SISSA/ISAS), Via
Bonomea 265, 34136 Trieste, Italy}

\pacs{75.47.Lx, 62.50.-p, 64.75.Qr}

\begin{abstract}The pressure dependence of the Curie temperature
T$_{C}(P)$ in La$_{0.75}$Ca$_{0.25}$MnO$_{3}$ was determined by
neutron diffraction up to 8 GPa, and compared with the
metallization temperature T$_{IM}(P)$ \cite{irprl}. The behavior
of the two temperatures appears similar over the whole pressure
range suggesting a key role of magnetic double exchange also in
the pressure regime where the superexchange interaction is
dominant. Coexistence of antiferromagnetic and ferromagnetic peaks
at high pressure and low temperature indicates a phase separated
regime which is well reproduced with a dynamical mean-field
calculation for a simplified model. A new P-T phase diagram has
been proposed on the basis of the whole set of experimental data.
\end{abstract}

\maketitle

Perovskite-like manganites La$_{1-x}$Ca$_{x}$MnO$_{3}$ (LaCa$x$)
exhibit a variety of physical properties depending on the Ca
concentration $x$. The strong correlation among magnetic,
electronic, orbital and transport properties of manganites makes
these systems particulary sensitive to external perturbations,
such as temperature variation, application of magnetic field or
high pressure (HP) \cite{dagotto,cheong}. The most popular
phenomenon is the Colossal Magneto-Resistance (CMR) \cite{chen}
observed over the $0.2<x<0.5$ range. Here, the system shows a
transition from a high temperature paramagnetic (PM) insulating
phase to a low temperature ferromagnetic (FM) metallic one and its
properties can be described by two competing mechanisms:
double-exchange (DE) \cite{zener} mechanism and Jahn-Teller (JT)
effect \cite{millis}. The DE model, which couples magnetic and
electronic degrees of freedom, qualitatively explains the observed
transition and the contiguity between the insulator to metal
transition temperature T$_{IM}$ and the Curie temperature T$_{C}$.
More realistic modeling is only obtained by introducing the JT
mechanism with its charge localizing effect \cite{millis}.

HP techniques were successfully employed to investigate the role
of JT structural distortion on manganite properties. The idea is
that lattice compression reduces JT distortion \cite{loa,ramprl},
increases hopping integral $t$, thus favoring the onset of
metallic phase \cite{irprl}. Several manganites show indeed an
increase of T$_{IM}$ on applying pressure up to 2-3 GPa
\cite{hwang,fontcuberta}. Nevertheless, experiments at higher
pressures cast doubts on the above scheme \cite{Cui3,Cui5}. Several HP experiments
were carried out on LaCa$x$ manganites. In particular, in
LaCa$0.25$ T$_{IM}$(P) was found to increase linearly only up to
P$\simeq$ 3 GPa, to bend down above 6 GPa and to approach an
asymptotic value close to room temperature \cite{irprl,prb}. An
antiferromagnetic (AFM) superexchange (SE) interaction between the
Mn magnetic moments \cite{feinberg}, which are formed by localized
t$_{2g}$ electrons, was proposed to be responsible for the
anomalous T$_{IM}(P)$ evolution at HP \cite{capone}. This peculiar
pressure behavior can be ascribed to a competition between the DE
and the SE interaction, which, is responsible for AFM order observed away from the CMR
concentration region at ambient pressure. Since the
strength of DE is proportional to the hopping integral $t$ and SE
is proportional to $t^{2}$, on increasing pressure (i.e. $t$) the
role of SE versus DE is continuously enhanced and it is expected
to become the dominant interaction in the HP regime
\cite{irprl,capone}. The pressure dependence of the magnetic
transition was recently investigated by neutron diffraction in
LaCa$0.33$ \cite{Kozenclo1} and LaCa$0.25$ \cite{Kozenclo2} over
the $0-4$ GPa range and an almost linear P-dependence of T$_{C}$
was observed. In LaCa$0.25$ rather close values of T$_{IM}$ and
T$_{C}$ \cite{Kozenclo2} were found over the whole P range. As a
confirm of the role of SE, evidence of AFM order coexisting within
the FM phase was also obtained at low temperature and HP
\cite{Kozenclo1,Kozenclo2}. The suppression of the long range FM
order at HP (around 23 GPa) in LaCa0.25 was deduced by X-ray
magnetic dichroism data \cite{yang}. The remarkable compression of
MnO$_{6}$ octahedra along the \emph{b} axis and the anomalous
change in the lattice strain at around 23 GPa together with the
monotonic decrease of the FM moment with pressure, suggested
indeed a transformation from an orbitally disordered FM phase to
an orbitally ordered AFM phase \cite{yang}. The coexistence of the
two magnetic phases in the pressure range between 2 and 23 GPa was
conjectured \cite{yang}. It is worth to note that although the
tendency toward mixed phases, whether structural or magnetic, was
recently observed in several manganites at ambient \cite{dagotto2}
and high pressure \cite{loa,prlarcan,bilayer,gallato}, the
theoretical investigation of this phenomenon was rather limited.

Although LaCa$0.25$ is the most extensively investigated manganite
system under pressure, no experimental data are available on
T$_{C}(P)$ over the P range where T$_{IM}$ is no more linear and
the SE mechanism is dominant over the DE. In the present paper we
extend the neutron diffraction measurements on LaCa$0.25$ up to 8
GPa in order to clarify the role and the relative strength of DE
and SE interactions and the coupling between T$_{C}$ and T$_{IM}$
in HP regime. The tendency towards phase separation is modeled by
using the simple theoretical approach successfully employed in
Ref. \cite{capone}.

The neutron diffraction experiment was carried out at the LLB
(Saclay, FR) using the G$6-1$ two-axis powder diffractometer with
a selected incident neutron wavelength $\lambda$ = 4.74 {\AA}. The
diffraction patterns were collected over the $1.5$-$300$ K
temperature range along several isobaric paths. Details about the
cell are reported in Ref. \cite{Goncha}. Sapphire anvil were used
at 2.1, 3.9, and 6.0 GPa and, diamond anvils were employed at 8 GPa. The sample volume ranged from several
millimeters (sapphire anvils) to $0.01$ mm$^{3}$ (diamond anvils).
The sample, prepared by a solid reaction method, was finely milled
and a mixture of sample and NaCl salt (1:2 volume proportion) was
loaded in the gasket hole together with small ruby spheres for
pressure calibration.

\begin{figure}
\includegraphics[width=8 cm]{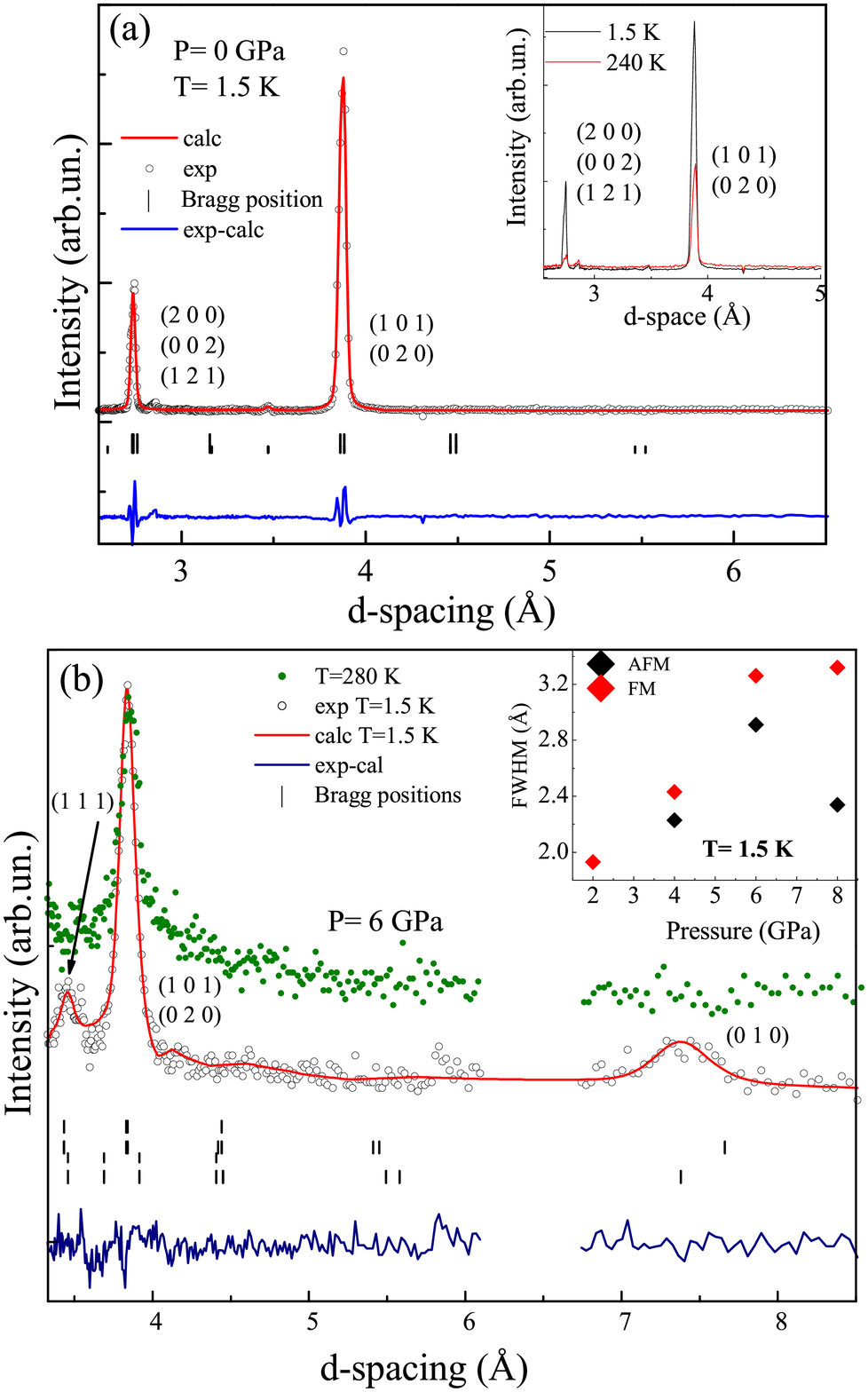}\\
\caption{(a): Diffraction pattern and Rietveld refinement at P=0
and T=1.5 K. Inset: comparison between the P=0 diffraction
patterns collected at T=240 and 1.5 K. (b) Diffraction patterns at
6 GPa collected at T=280 and 1.5 K. Peaks at $d$ = 3.47 and $d$ =
7.38 {\AA} are ascribed to AFM order. Calculated positions of the
Bragg reflections are also shown: the two upper sets correspond to
the nuclear and magnetic phase in agreement with
Ref.\cite{Kozenclo2,yang}, the two lower sets are associated to
the second structural and magnetic phase. Inset: FWHM of AFM and
nuclear peaks versus pressure at T=1.5 K.} \label{fig1}
\end{figure}

Ambient pressure data were collected over a wide $d$-space range
2.5 {\AA} $<d<$ 54 {\AA} (2$\theta=5-140^{\circ}$) using two
detector positions. Diffraction patterns at high and low
temperature are shown in the inset of Fig.\ref{fig1}(a).
At low temperature, the strong increase of the peak intensities at
d=3.87 $\AA$ ((1 0 1)/(0 2 0)) and d=2.74 $\AA$ ((2 0 0 )(0 0 2)(1
2 1)) can be ascribed to the FM contribution \cite{Kozenclo2}
(inset of Fig.\ref{fig1}(a)). The diffraction pattern was thus
refined using the Pnma orthorhombic structure with an associated
FM order. The lattice parameters (\emph{a}=5.47 $\AA$,
\emph{b}=7.72 $\AA$, \emph{c}=5.50 $\AA$) and the magnetic moment
$\mu=3.57 \mu_B$ obtained at ambient pressure by Rietveld refinement are in good
agreement with the results of Ref. \cite{radaelli}.

In order to reduce the acquisition time at HP, we focused on the main peak at d=3.87$\AA$ and diffraction
patterns were collected exploiting only one detector position
(2$\theta=12-91^{\circ}$ and 3.3 $\AA$ $<d<$ 22 $\AA$). Data collected at 6 GPa are shown in
Fig.\ref{fig1}(b) where new magnetic reflections are found at
d=7.38 $\AA$ (0 1 0) and d=3.47 $\AA$ (1 1 1). These peaks are unambiguously
observed at P$=3.9$ GPa and upwards, whereas the data collected at P$=2.1$ does not allow any definitive conclusion. Nevertheless, a
new pressure induced peak (d=7.47 $\AA$ at 1.5 GPa) was previously
reported in Ref.\cite{Kozenclo2} and ascribed to an AFM order with
A-type structure and propagation vector $k=(0 0 0)$. This suggests
that the AFM peak is probably too weak compared with the
signal/noise ratio to be observed in the patten collected at
P$=2.1$ GPa.

Looking at the present data, a discrepancy is observed between the observed AFM peak position and the expected
\emph{d}-spacing calculated on the basis of the lattice parameters.
For example, considering the lattice parameters ((\emph{a}=5.41$\AA$, \emph{b}=7.66$\AA$  \emph{c}= 5.45$\AA$).) obtained at 6 GPa, the (0 1 0) AFM peak is expected to be found at 7.66 $\AA$, whereas it is centered at d=7.38 $\AA$ (see Fig.\ref{fig1} (b)).
At this point, it is worth noticing that both Kozenklo and Ding observed an uniaxial anisotropic compression of the MnO$_{6}$ octahedra along the \emph{b} axis of the orthorhombic structure, which is consistent with this shift of the AFM peak position \cite{Kozenclo2, yang}.
These findings suggest that the two magnetic phases, the AFM and the FM, are not associated to the same nuclear phase.
Two possible explanations can be proposed: i) the AFM order is not commensurate with the nuclear phase, as already reported for some manganite compounds
\cite{landsgesell}; ii) the AFM order is associated to a nuclear
phase with the same Pnma symmetry but different values of lattice
parameters.

In support of the ii) hypothesis, the positions of nuclear and
magnetic reflections were obtained above 4 GPa using Fullprof
software (see Fig.\ref{fig1}(b)) \cite{fullprof}. For example at 6
GPa, two Pnma nuclear structures and two magnetic orders were
considered: the first one associated to FM order with the set of
lattice parameters reported in Ref. \cite{Kozenclo2, yang}) and
the second one associated to an AFM order with $\emph{a}$=5.58
$\AA$, \emph{b}=7.38 $\AA$ and \emph{c}=5.49 $\AA$. The AFM unit
cell was described by four Mn atoms: Mn$_{1}=$$(0 0 \frac{1}{2})$,
Mn$_{2}=$ $(\frac{1}{2} 0 0)$ Mn$_{3}=$ $(0 \frac{1}{2}
\frac{1}{2})$ Mn$_{4}=$ $(\frac{1}{2} \frac{1}{2} 0)$. The
magnetic structure is a combination of the $\Gamma_{5}$ and
$\Gamma_{7}$ representations. Bragg reflections calculated for the
AFM structure (A-type) using the R$_{x}=(++--)$ and R$_{y}=(++--)$
sequence of magnetic moments and propagation vector $k=(0 0 0)$,
were well consistent with the experiment as shown in
Fig.\ref{fig1}(b). The magnetic moments lay on the \emph{ab} plane
and they are placed in a way that is similar to that one observed
in LaMnO$_{3}$ at high pressure \cite{gaudart}. The fractional
volumes of the FM and the AFM phases were obtained for the neutron
pattern collected at T= 5K and were found to pass from 87.9 $\%$
(FM) and 12.1 $\%$ (AFM) at 4 GPa to 76.3 $\%$ (FM) and 23 $\%$
(AFM) at 8 GPa. The FWHM of the FM $(1 0 1)/(0 2 0)$, and the AFM
peak $(0 1 0)$, show opposite behaviors on increasing the
pressure: the former increases while the latter decreases (see the
inset of Fig.\ref{fig1}(b)). This behavior is thus consistent with
the hypothesis ii), that is lattice compression simultaneously
induces both AFM order and a separation between two isostructural
nuclear phases.

The emerging scenario is consistent with a founded
physical picture. If AFM order is driven by SE interaction, Mn
orbitals are expected to order in a way that reduces \emph{b}
axis. The connection between lattice and orbital degrees of
freedom, as investigated in several orthorhombic manganites
\cite{bilayer,woodward, gallato2}, suggests that the development
of orbital ordering results in a contraction of \emph{b}
parameter, and if $\emph{b}/\sqrt{2}< \emph{c}< \emph{a}$ the
occurrence of an orbital ordering can be conjectured
\cite{woodward,gallato2}.
\begin{figure}
\includegraphics[width=8cm]{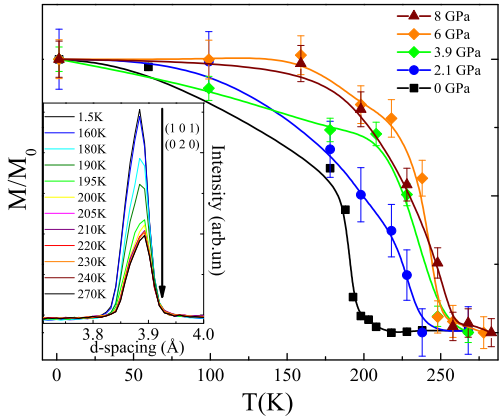}\\
\caption{Normalized magnetic moments versus temperature (solid
lines are guides to the eye). Inset: temperature dependence of the
FM peak {h,k,l}={1,0,1}} \label{FIG2}
\end{figure}
On the other hand, if $\emph{c}> \emph{a}\simeq
\emph{b}/\sqrt{2}$, orbital disorder is expected. It appears thus
reasonable to associate FM and AFM order to orthorhombic phases
with large and small \emph{b} parameters: 7.66 $\AA$ and 7.38
$\AA$ respectively.

The FM to PM transition can be identified looking at the
temperature dependence of the integrated intensity of the main
peak (d=3.87$\AA$) (inset in Fig. \ref{FIG2}). The results of the refinements do not considerably change including or excluding the Debye-Waller factor suggesting that the temperature dependence of the nuclear intensities can be neglected.
Therefore, the Debye-Waller factor was neglected over the scattering vector range relevant to
magnetic peaks. After proper background correction, the integrated
peak intensities, subtracted by the nuclear contribution, can be
reasonably considered proportional to the effective magnetic
moments M. The temperature dependence of the normalized magnetic
moments M/M$_{0}$ (where M$_{0}$ is the effective magnetic moment
at the lowest T) is shown in Fig.\ref{FIG2} at different
pressures.

A rapid increase of T$_{C}$ on applying pressure is evident over
the low pressure range (0-4 GPa), whereas a strongly reduced
pressure dependence is observed above 4 GPa. We notice that the observed broadening of the transition can
be likely ascribed to strain effects due to pressure gradients.
The data were analyzed using a modified Brillouin function to
describe the temperature dependence of the magnetization,
according to the Weiss molecular field theory \cite{brillouin}. To
avoid the intrinsic ambiguity of the definition of both T$_{C}$
and T$_{IM}$, we used the temperature variations $\Delta$T from
ambient pressure T$_{C}$(P=0) and T$_{IM}$(P=0) to compare the
pressure dependence of T$_{C}$(P) and T$_{IM}$(P) \cite{irprl}.
\begin{figure}
\includegraphics[width=8cm]{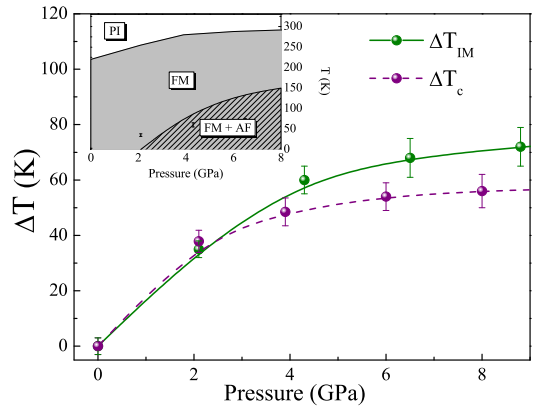}\\
\caption{Temperature variation of T$_{C}$ and T$_{IM}$
\cite{irprl} versus pressure. Inset: tentative P-T phase diagram. The onset of the AFM peak at 2 GPa refers to Kozenklo \emph{et al.} work \cite{Kozenclo2}}
\label{fig3}
\end{figure}
As shown in Fig. \ref{fig3}, the pressure dependence of $\Delta$T
is basically the same for both T$_{C}$ and T$_{IM}$: a linear
increase at low pressure followed by a saturation towards an
asymptotic value, above 6 GPa. This result further confirms the
onset of a pressure driven localizing mechanism that, becoming
competitive with DE delocalizing mechanism at high pressure,
strongly affects both the transition temperatures T$_{C}$ and T$_{IM}$. Moreover, the onset of an AFM state at low temperature
coexisting with the FM order confirms the SE-AFM nature of the
localizing interaction. This suggests a phase separation scenario
at low temperature, as depicted in the inset of Fig. \ref{fig3}
where a tentative P-T phase diagram is shown. We finally notice
that $\Delta$T$_{C}(P)$ slightly deviates from $\Delta$T$_{IM}(P)$
only at very high pressure. This means that decreasing temperature
at a pressure above $\sim$ 4 GPa, LaCa$0.25$ enters a metallic
phase before achieving the FM state. This features might be
consistent with a percolation scenario among metallic clusters in
a PM insulating bulk, as previously suggested by HP optical
measurements \cite{prlarcan}.

In order to understand theoretically if the pressure-induced increase of AFM-SE
can, in turn, induce a phase separation, we considered a
simplified model for manganites similar to that used in Refs. \cite{caponeciuchi,capone}.

\begin{eqnarray} 
\label{hamiltonian}
H&=&-t \sum_{\langle ij\rangle\sigma} {\left( c_{i,\sigma}^{\dagger} c_{j,\sigma} c_{j,\sigma}^{\dagger} c_{i,\sigma} \right)} -\mu \sum_{i\sigma} c^{\dagger}_{i\sigma}c_{i\sigma}+ \nonumber \\
&& - J_H \sum_{i} {\vec{\sigma}_i \cdot \vec{S}_i}+  \nonumber \\
&& + J_1 \sum_{\langle ij\rangle}\vec{S}_i \cdot \vec{S}_j - g \sum_i \left( n_i \right) (a_i+a^{\dagger}_i ) + \omega_0 a^{\dagger}a.
\end{eqnarray}

A single band of itinerant electrons with nearest-neighbor hopping $t$ replaces the two $e_g$ bands, and a Holstein coupling to a local Einstein phonon of frequency $\omega_0$ mimics the JT interaction.
The DE physics is introduced through an FM coupling between the itinerant band and localized spins $\vec{S}_i$ associated to the $t_{2g}$ orbitals.
The localized spin are, in turn, antiferromagnetically coupled with a coupling constant which is small at ambient pressure and grows with increasing pressure ad discussed in Ref. \cite{capone}.
We work in the grandcanonical ensemble, where the chemical potential $\mu$ determines the particle density in the $e_g$ bands.
The density $n = \langle \sum_{i\sigma} n_{i,\sigma}\rangle$, where $n_{i,\sigma} = c_{i,\sigma}^{\dagger}c_{i,\sigma}$ is related to the Ca doping by the relation $x = 1-n$

The study of the thermodynamic instability associated to phase
separation requires to work in the thermodynamic limit, which can
be addressed using dynamical mean-field theory
(DMFT)\cite{revdmft}. We studied the Holstein-DE model
(\ref{hamiltonian}) at T=0 using DMFT and exact diagonalization to
solve the numerical part of DMFT. We used our implementation of
DMFT, as discussed in Ref. \cite{cdg_ed}, with exact
diagonalization using 8 levels in the discretized bath.

Since the aim of this calculation is to understand if, on general grounds, phase separation can be expected from the competition between AFM and FM, we introduced some safe approximations.
In particular we treated the localize $t_{2g}$ spins as classical variables neglecting quantum fluctuations.
Moreover, in light of the experimental suggestions, only the two configurations FM and commensurate AFM ordering were considered, neglecting intermediate orientations and more exotic spin patterns.
We computed the total energy for these two phases and for the nonmagnetic phase as a function of the Ca concentration.
The effect of pressure has been included in the calculations adopting the parameters of Ref. \cite{capone}, with the SE coupling scaling as the square of the hopping parameter.
Even though our model is too simplified to attempt a quantitative comparison with the experimental phase diagram, it is expected to properly reproduce the competition between magnetic phase which ultimately leads to phase separation.

\begin{figure}
\includegraphics[width=7 cm]{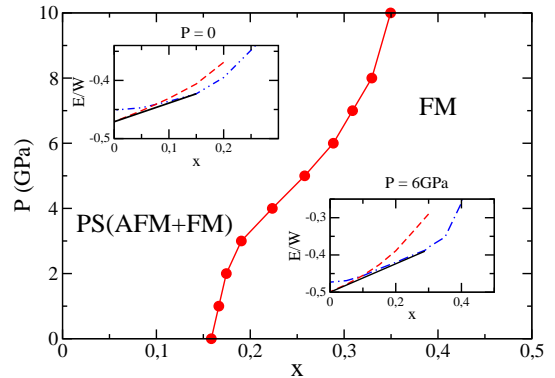}
\caption{P-x phase diagram for the Holstein-DE model with parameters from Ref.\cite{capone}.
A pressure dependence line separates a phase separation region from a homogeneous ferromagnetic state. Phase separation is favored by increasing pressure.
Insets: the x-dependence of the Free Energy normalized to the bandwidth at P=0 and 6 GPa. Dashed red (blue) line represents
the x-dependence for the AFM (FM) phase.
Solid black line is a Maxwell construction whose boundaries determine the doping range for which phase separation takes place at the chosen pressure.}\label{fig4}
\end{figure}

Phase separation is associated to a negative charge compressibility $\kappa = \partial n/\partial \mu$, which corresponds to a negative curvature of the energy as a function of density (or equivalently on doping).
In order to detect phase separation we need to solve the system for various values of Ca concentration and analyze the curvature of the ground state energy around the experimentally relevant case $x=0.25$.
In our case each of the independent solutions (FM, AFM, non magnetic) has the correct positive curvature for every value of $x$ (see curves in the two inset of Fig. \ref{fig4}).
However, when two curves cross as a function of doping and the system undergoes a first-order transition from one phase to the other, the energy has a cusp, and, more importantly, one can lower the energy by dividing the system into a fraction of AFM and a fraction of FM.
This is realized through a Maxwell construction (solid black line in the two panels) and physically corresponds precisely to phase separation.

In Fig.\ref{fig4} we show the pressure vs Ca doping phase diagram obtained with DMFT.

At zero pressure we find phase separation between AFM and FM only close to $x=0$
\cite{jarrell}. Our simplified model shows indeed phase separation below $x \simeq 0.16$, while larger doping correspond to homogeneous FM phases.
Increasing the pressure increases the phase separated region, pushing its boundary to larger doping. At $P \simeq 5$ GPa the phase separation region reaches $x=0.25$.
Further increasing pressure at fixed $x$ enhances the size of the AFM fraction of the system. Despite the simplifications introduced in our model, the characteristic pressure leading to phase separation is reproduced accurately by the theoretical calculations, confirming that the enhancement of the AFM superexchange can indeed account for the observed pressure-driven phase separation.

In summary, the pressure dependence of the magnetic temperature
T$_{C}(P)$ was obtained up to 8 GPa and found to be rather close
to $T_{IM}(P)$, suggesting that DE model still holds at HP. The
reinforced role of SE localizing mechanism, competing with
delocalizing DE, was confirmed by $T_{C}(P)$ and by the onset of
an AFM state at low temperature. The novel antiferromagnetic
structure is not related to the crystalline structure associated
to the FM phase and observed at ambient pressure. The data are
consistent with a picture in which lattice compression
simultaneously induces both AFM order and a separation between two
isostructural nuclear phases. The simultaneous appearance of FM
and AFM peaks depicts a phase separation scenario promoted by
pressure. This tendency towards a phase separation is properly
reproduced, using a simplified theoretical model
indicating that formation of domains is promoted by pressure.\\

\acknowledgments M.C. is funded by the European Research Council
under FP7/ERC Starting Independent Research Grant "SUPERBAD"
(Grant Agreement n. 240524) and MIUR PRIN 2007 Prot.
2007FW3MJX003.


\begin{thebibliography}{0}
\bibitem{nota32} Details of the calculations and a comparison with
more realistic models of the manganites (inclusion for orbital
degeneracy of the $e_{g}$ electrons and proper JT interaction,
quantum description of $t_{2g}$ spins) will be presented in a
forthcoming work.
\end{thebibliography}
\end{document}